\documentclass[preprint]{aastex}

\usepackage{graphicx}
\usepackage{rotating}
\usepackage{amsmath}
\usepackage{natbib}
\usepackage{subfigure}

\begin{document}

\title{High energy emission components in the short GRB 090510}

\author{Alessandra Corsi\altaffilmark{1,2,3}, Dafne Guetta\altaffilmark{4} and Luigi Piro\altaffilmark{2}}

\email{alessandra.corsi@roma1.infn.it}
\email{guetta@oa-roma.inaf.it}
\email{luigi.piro@iasf-roma.inaf.it}

\altaffiltext{1}{Universit\`a degli studi di Roma ``Sapienza'' and INFN-Sezione di Roma, Piazzale Aldo Moro 2, 00185 Roma, Italy.}
\altaffiltext{2}{INAF - Istituto di Astrofisica Spaziale e Fisica Cosmica di Roma, Via Fosso del Cavaliere 100, 00133 Roma, Italy.}
\altaffiltext{3}{Current address: LIGO Laboratory, California Institute of Technology, Pasadena, CA 91125, USA. E-mail: corsi@caltech.edu} 
\altaffiltext{4}{INAF - Osservatorio Astronomico di Roma, Via Frascati 33, 00040 Monte Porzio Catone, Italy.}

\begin{abstract}We investigate the origin of the prompt and 
delayed emission observed in the short GRB 090510. We use the 
broad-band data to test whether the most popular theoretical 
models for gamma-ray burst emission can accommodate the 
observations for this burst. 
We first attempt to explain the soft-to-hard spectral evolution 
associated with the delayed onset of a GeV tail with the 
hypothesis that the prompt burst and the high energy tail both 
originate from a single process, namely synchrotron emission 
from internal shocks. Considerations on the compactness of the 
source imply that the high-energy tail should be produced in a 
late-emitted shell, characterized by a Lorentz factor greater 
than the one generating the prompt burst. However, in this 
hypothesis, the predicted evolution of the synchrotron peak 
frequency does not agree with the observed soft-to-hard 
evolution. Given the difficulties of a single-mechanism 
hypothesis, we test two alternative double-component scenarios. 
In the first, the prompt burst is explained as synchrotron 
radiation from internal shocks, and the high energy emission 
(up to about 1 s following the trigger) as internal shock 
synchrotron-self-Compton. In the second scenario, in view of 
its long duration ($\sim 100$ s), the high energy tail is 
decoupled from the prompt burst and has an external shock 
origin. In this case, we show that a reasonable choice of 
parameters does indeed exist to accommodate the optical-to-GeV 
data, provided the Lorentz factor of the shocked shell is 
sufficiently high. Finally, we attempt to explain the chromatic 
break observed around $\sim 10^3$ s with a structured jet 
model.  We find that this might be a viable explanation, and 
that it lowers the high value of the burst energy 
derived assuming isotropy, $\sim 10^{53}$ erg, below
$\sim 10^{49}$ erg, more compatible with the energetics from a 
binary merger progenitor. \end{abstract}


\shortauthors{A. Corsi, D. Guetta, L. Piro}

\shorttitle{High energy emission in GRB090510}

\maketitle

\section{Introduction}
Traditionally divided in ``long'' and ``short'' on the basis of 
their $\gamma$-ray duration \citep[longer or shorter than 2 s,][]{Kouvelioutou1993}, 
gamma-ray bursts (GRBs) are characterized by a prompt release of 
$\gamma$- and X-ray photons, followed by a multi-wavelength 
afterglow \citep{Costa1997} emission. The fireball model 
\citep[e.g.][]{Meszaros1992,Sari1998} explains the GRB 
electromagnetic emission as a result of shock dissipation in a 
relativistic flow, or ``fireball'', taking place at distances greater 
than $10^{-5}$-$10^{-2}$ pc from the central source. Despite such a 
large distance, electromagnetic observations have revealed important 
clues about the progenitors, favoring two main models: 
the coalescence of a binary consisting of two neutron stars or a 
neutron star and a black hole for short GRBs; and the death of 
massive stars (collapsars) for long GRBs.
Both of these are commonly believed to end in a BH-plus-torus system, 
where torus accretion powers the fireball jet.  

In the internal-external shock scenario of the fireball model 
\citep[see e.g.][]{Meszaros1993,Sari1998}, GRB prompt and afterglow 
emissions are thought to be produced by particles accelerated via 
shocks in an ultra-relativistic outflow released during the burst 
explosion. While the prompt emission is related to shocks developing 
in the ejecta (internal shocks, IS), the afterglow arises from the 
forward external shock (ES) propagating into the interstellar medium (ISM).
Synchrotron and synchrotron-self-Compton (SSC) emission by the 
accelerated electrons are typically invoked as the main radiation 
mechanisms.
The Fermi satellite\footnote{http:$//$fermi.gsfc.nasa.gov} is 
currently bringing exciting new results, detecting high energy 
($\sim$ GeV) extended tails whose presence is particularly 
intriguing in the case of short GRBs 
\citep[e.g.][]{Abdo2009,Giuliani2009,GCN8407,GCN9334}. 
Fermi observations of GRB 081024B and GRB 090510 clearly point to 
the existence of a longer-lasting high energy tail in the GeV range 
following the main event, motivating a deeper exploration and 
re-examination of the fireball physics and radiative mechanisms 
\citep[e.g.][]{Asano2009,Corsi2009,Gao2009,Kumar2009,Zou2008}.

Here we study the conditions under which the high energy 
observations of GRB 090510 can be accommodated within the most 
popular theoretical models. This work is organized as follows. In 
Sec.~\ref{sec1} we summarize the observations for GRB 090510. In 
Sec.~\ref{sec2} we discuss in more detail the spectral properties of 
the emission during the first second, and their implications for the 
compactness of the source. We test whether the complex emission 
observed during the first 1 s can be attributed to a single 
component, specifically synchrotron emission from IS. After showing 
that a single mechanism does not offer a straightforward 
explanation, we test an IS synchrotron plus SSC scenario. In 
Sec.~\ref{sec5} we examine separately the high energy emission in 
the context of the ES scenario. Finally, in Sec.~\ref{sec6} we give 
our conclusions. Hereafter we adopt as $T_0$ the onset of the main 
GRB pulse which, as specified in the following section, is about 
$\sim 0.5$ s after the precursor that triggered the Fermi/GBM. Also, 
hereafter $\epsilon_e$ and $\epsilon_B$ are the fraction of energy 
going into electrons and magnetic fields, respectively; $n$ is the 
ISM density in particles/cm$^3$;  $E_{iso}$ is the isotropic kinetic 
energy of the fireball; and $p$ is the power-law index of the 
electron energy distribution in the shock.


\section{The Observations}
\label{sec1}

Characterized by a $T_{90}$ of 0.3 s \citep{GCN9337}, which places 
GRB 090510 in the short GRB category, the main burst was followed by 
an extended high energy tail, observed by both AGILE/GRID and 
Fermi/LAT on a timescale much longer than the prompt burst event 
\citep{Abdo2009,Giuliani2009}.
AGILE detected GRB 090510 at $T_0=00:23:00.5$ UT, May 10, triggering 
on the sharp main peak of the GRB (about 0.5 s after the Fermi 
trigger on a smaller precursor). The Swift/BAT also triggered around 
the main burst peak \citep[][and references therein]{Hoversten2009} 
at $00:23:00.4$ UT. Hereafter we adopt as $T_0$ the onset of the 
main peak.  
The 0.3-10 MeV and $\gtrsim 25$ MeV emission of the burst showed a 
clear dichotomy between the low- and high-energy $\gamma$-ray 
emissions, so that two time intervals were defined: interval I from 
$T_0$ to $T_0+0.2$ s, and interval II from $T_0+0.2$ s to $T_0+1.2$ s 
\citep{Giuliani2009}. The main peak in the AGILE calorimeter (MCAL, 
0.35-100 MeV) ended around 0.2 s, at which time the signal suddenly 
started to be observed in the GRID ($> 100$ MeV). The Swift/BAT 
light curve showed two pulses between 0.2 s and 0.3 s, of amplitude 
much smaller than the first peak \citep{GCN9337}. 

The AGILE photon spectrum of Interval I is well modeled by a 
power law with index $\alpha=-0.65^{+0.28}_{-0.32}$ and exponential 
cutoff $E_c=2.8$ MeV \citep[see the top panel of Fig.~4 in][]{Giuliani2009}. 
This is consistent with the Fermi observations, which are well 
modeled by a Band spectrum \citep{Band1993} with $E_{peak}=2.8$ MeV, 
$\alpha=-0.59\pm0.04$, $\beta<-5$, and a normalization constant 
$A\sim 8\times10^{-2}$ ${\rm ph~cm}^{-2}{~\rm s}^{-1}{~\rm keV}^{-1}$ 
\citep{Abdo2009}. 

During interval II, the spectrum undergoes a soft-to-hard evolution. 
Specifically, for AGILE \citep[see the lower panel of Fig.~4 in][]{Giuliani2009}, 
the MCAL spectrum is a power-law, $N(E) \sim C (E/100 {\rm~ keV})^{\beta}$, 
of photon index $\beta=-1.58^{+0.13}_{-0.11}$, and 
$C=1.6 \times 10^{-2}~{\rm ph~cm}^{-2}{~\rm s}^{-1}{~\rm keV}^{-1}$ 
(derived by considering that the emitted 0.5 MeV - 10 MeV fluence 
during interval II was of $\sim 3.1\times10^{-6}$ erg cm$^{-2}$; see 
\citet{Giuliani2009}). This spectrum is consistent with the one 
measured by the GRID, which is well fit by a power law with index 
$\beta=-1.4\pm0.4$, for a 25-500 MeV fluence of 
$\sim 2.12\times10^{-5}$ erg cm$^{-2}$ \citep{Giuliani2009}. 

Time-resolved spectral fits of Fermi data from 
interval II show a progressive evolution from a Band plus power-law 
to a single power-law spectrum. Between $T_0+0.1$ s and $T_0+0.3$ s, 
the Band component is still evident and has best-fit spectral 
indices of $\alpha=-0.48\pm0.07$, $\beta=-3.09^{+0.21}_{-0.35}$. 
In the last two temporal bins the contribution of the Band component 
becomes less important, with the peak flux decreasing by about an 
order of magnitude.  The power-law component in these bins has a 
best-fit photon index/normalization at 1 GeV of 
$\beta=-1.54^{+0.07}_{-0.04}$ / $A_{pow}=(6.4^{+1.6}_{-1.2})\times10^{-9}$ 
${\rm ph~cm}^{-2}{~\rm s}^{-1}{~\rm keV}^{-1}$, and 
$\beta=-1.92^{+0.20}_{-0.22}$ / $A_{pow}=(3.7^{+1.3}_{-1.1})\times10^{-9}$ 
${\rm ph~cm}^{-2}{~\rm s}^{-1}{~\rm keV}^{-1}$, 
respectively \citep{Abdo2009}.  

At a redshift of $z=0.903$ \citep{GCN9353}, the 10 keV - 30 GeV 
measured burst fluence during $\sim 0.5$ s since $T_0$ implies an 
isotropic energy release of $\sim 10^{53}$ erg \citep{Abdo2009}, 
which is extremely high for a short GRB.

A temporal analysis of the high-energy tail at energies above 0.1 
GeV as observed by the
Fermi/LAT shows a GeV flux rising in time as $\sim t_{obs}^2$, and 
decaying as $\sim t_{obs}^{-1.5}$ up to $\sim 200$ s following the 
trigger \citep{Ghirlanda2009}. Similarly, the signal detected by the 
AGILE/GRID showed that during Interval II and later on, up to 10 s 
after $T_0$, the emission can be described by a power-law temporal 
decay of index $\delta=-1.30\pm0.15$ \citep[see the top panel of 
Fig.~3 in][]{Giuliani2009}.

The Swift XRT began observing the burst about 100 s after the 
trigger \citep{Hoversten2009}. In X-rays, a steepening is 
observed around $1500$ s, which changes the power-law temporal 
decay index from $\delta = -0.74\pm0.03$ to $\delta=-2.18 \pm 0.10$. 
The optical light curve first rises until $\sim 1600$ s, and 
then decreases as a power law with 
$\delta=-1.13^{+0.17}_{-0.09}$ \citep{DePasquale2009}. 
This power-law decay is shallower than the one observed in X-rays 
after about $1500$ s. We also note that the optical emission is 
observed to peak much later than the extended high energy tail 
observed by the LAT.

\section{The first 1 s of emission}
\label{sec2}
The dichotomy of the spectral behavior observed during the first 
second of emission suggests that the properties of the source are 
evolving between interval I and interval II. While during 
interval II the observation of a $\sim 30$ GeV photon requires an 
optically thin source in the GeV range, during interval I the 
absence of emission above 100 MeV and the unusually steep high 
energy photon index ($\beta<-5$, much smaller than values typically 
observed in GRB prompt spectra), suggest that thickness due to 
pair production plays a role.

The key parameter determining the optical thickness due to pair 
production is the Lorentz factor of the shell. As we 
show in detail in this section, the bigger the Lorentz factor, 
the lower the optical thickness. Thus, a scenario possibly 
explaining GRB 090510 observations could be the following. The 
central engine emits a first shell with Lorentz factor $\Gamma_{I}$, 
responsible for the first peak observed in the GRB light curve, 
which covers Interval I. This peak is characterized by an observed 
spectrum with no emission above 100 MeV and an extremely steep high 
energy spectral slope, thus suggesting that $\Gamma_{I}$ is such 
that the source is optically thick above 100 MeV. Later on, the 
central engine emits a series of shells responsible for the other 
multiple peaks observed during interval II. These shells are 
characterized by a Lorentz factor between $\Gamma_{I}$ and 
$\Gamma_{II}$, where $\Gamma_{II}>\Gamma_{I}$ and is such that the 
source is transparent to GeV photons. From the point of view of the 
physical properties of the source, this implies that the GRB central 
engine should be emitting shells with progressively higher 
velocities. The hypothesis that the source emits shells of different 
velocities is indeed the basis of the IS model. 

In the above scenario, time-resolved spectroscopy during interval II 
should show a progressive transition from an optically thick to an 
optically thin spectrum in the GeV range. Also, any spectrum 
obtained by integrating over multiple peaks happening during 
interval II would show a superposition of spectra emitted by shells 
with different $\Gamma$ factors, progressively more transparent to 
GeV photons. Time-resolved spectroscopy by Fermi during interval II 
does indeed show a transition from a spectrum peaking around few 
MeV (Band component) to one with substantial emission in the GeV 
range (power-law component). Also, these components are observed 
simultaneously in the spectrum integrated from $T_0 + 0.1$ s 
to $T_{0}+0.3$ s \citep[red curve in Fig.~2 of][]{Abdo2009}, which 
includes at least two peaks (see Fig.~1 in \citet{Abdo2009})
following the main one. Thus, on general lines, the 
observed spectral evolution during the first 1 s of emission from 
GRB 090510 is consistent with the hypothesis of a transition from 
an optically thick to an optically thin spectrum in the GeV range, 
which would naturally explain the delayed onset of the GeV tail 
observed by the LAT. 

To validate the scenario outlined above, however, a more 
quantitative test is necessary. It is worth stressing that, 
according to the observations, between interval I and II not only is 
the source becoming optically thin to GeV photons, but also the 
spectral shape of the observed emission is changing substantially. 
In particular, a crucial point to verify is whether the evolution of 
the Lorenz factor required to justify a transition toward a smaller 
thickness in the GeV range also agrees, within the IS model, with a 
shift in the spectral peak from a few MeV to more than 1 GeV 
(as observed in the transition from a Band to a power-law spectrum). 
In what follows, we analyze this scenario in detail.

\subsection{Interval I}
\label{sec3}
\subsubsection{Thickness to pair production}
\label{tick}
We can argue that during interval I, and in particular between 
$T_0$ and $T_0+0.1$ s \citep[see][]{Abdo2009}, the unusually steep 
high-energy photon index observed by Fermi ($\beta<-5$) is due to 
optical thickness from pair production on an underlying Band 
function with $A$, $E_{peak}$ and $\alpha$ as the observed ones 
(see Sec.~\ref{sec1}), but with $\beta\gtrsim-5$. More specifically, 
hereafter we make the hypothesis that the true (unabsorbed) high 
energy spectral index has a value of $\beta=-3.5$. In fact, 
according to the complete spectral catalog of BATSE bright GRBs 
\citep{Kaneko2006}, the tail of the distribution for the spectral 
index of well-modeled spectra is around $\beta=-3.5$. 
While the BATSE catalog did not show any significant difference 
between the spectral parameters of short GRBs and those of long 
ones, we note that two of the short GRBs in that sample indeed had 
$\beta \sim -3.5$ \citep[see Table 14 in][]{Kaneko2006}. 
Moreover, being in the tail of the distribution, $\beta=-3.5$ would 
reconcile GRB 090510 observations with the more commonly observed 
properties of GRB prompt spectra, while minimizing the implied value 
of the $\tau_{\gamma\gamma}$ for pair production. In fact, we cannot have 
$\tau_{\gamma\gamma}>>1$ if the observed spectrum is non-thermal and 
the light curve shows high temporal variability. We also note that 
an unabsorbed $\beta$ of $-3.5$ would be consistent, within the 
errors, with $\beta=-3.09^{+0.21}_{-0.35}$ of the Band component 
observed by Fermi \citep{Abdo2009} during $T_0+0.1$ s and $T_0+0.2$ 
when, as discussed in Sec.~\ref{sec2}, we expect a contribution from 
less thick shells whose spectrum (in the observed energy band) is 
evolving from a Band to a power-law shape. This of course should be 
taken with the caveat that even $\beta=-3.09^{+0.21}_{-0.35}$ could 
still be affected by absorption.

The high-energy part of a Band spectrum reads
\begin{equation}
	N(E)=0.08 \left(\frac{(\alpha-\beta)E_{peak}}{e(2+\alpha)100 {\rm~ keV}}\right)^{\alpha-\beta}\left(\frac{E}{100 {\rm~ keV}}\right)^{\beta} \frac{\rm ph}{\rm cm^{2}s~keV}=C_{Band}\left(\frac{E}{100 {\rm~ keV}}\right)^{\beta}\frac{\rm ph}{\rm cm^{2}s~keV} \, ,
\end{equation} 
with $\alpha\sim -0.59$, $E_{peak}\sim 2.8$ MeV for GRB 090510 (see 
Sec.~\ref{sec1}). This equation is obtained from Eq.~(1) of 
\citet{Band1993} by using $E_{peak}=(2+\alpha)E_0$ 
\citep[see e.g.][]{Piran1999}. Note also that the multiplicative factor $e^{\beta-\alpha}$ 
in Eq.~(1) of \citet{Band1993} is included in the first 
factor in parenthesis in our above equation. If 
the \textit{true} spectrum has $\beta=-3.5$, then 
\begin{equation}
	C_{Band}=0.08 \left(\frac{(-0.59+3.5)2.8 {\rm~ MeV}}{e(2-0.59)100 {\rm~ keV}}\right)^{-0.59+3.5}\sim584 \, ,
\end{equation}
and we should have 
\begin{equation}
	\tau_{\gamma\gamma}(100 {\rm~ MeV})\gtrsim 4 
	\label{reqtau}
\end{equation}
to reconcile this spectrum with the observed $\beta\lesssim -5$. 
The $\tau_{\gamma\gamma}$ for pair production is expressed as 
follows \citep{Lithwick2003}:
\begin{equation}
	\tau_{\gamma\gamma} (E)\sim \frac{0.1 \sigma_T N_{\gamma>E_{an}(E)}}{4\pi R^2} \, .
\label{tau}
\end{equation}
In the above relation, $\sigma_T$ is the Thompson cross section, $R$ 
is the size of the source, and $N_{\gamma>E_{an}(E)}$ is the number 
of target photons; i.e., the number of photons with energy above 
$E_{an}$, where
\begin{equation}
	E_{an}(E)=\frac{(\Gamma m_e c^2)^2}{E (1+z)^2}=\frac{2.6\times10^{5}\Gamma^{2}}{(E/{\rm keV})(1+z)^2} \, {\rm keV} \, .
	\label{ean}
\end{equation}
This accounts for the fact that a photon with energy $E$ in the 
observer frame may be attenuated by pair production through 
interaction with softer photons, whose energy (also in the observer 
frame) is equal to or greater than $E_{an}(E)$. Also, for a power-
law spectrum of the form 
\begin{equation}
	N(E)=C (E/100 {\rm~ keV})^{\beta} \frac{\rm ph}{\rm cm^{2}s~keV}
\end{equation} 
one has
\begin{eqnarray}
	N_{\gamma>E_{an}(E)}=\frac{C ~4 \pi (d_L/{\rm cm})^{2}(\delta t_{obs}/{\rm s}) (E_{an}(E)/{\rm keV})^{1+\beta}}{-(1+\beta)(100)^{\beta}(1+z)^{2}} \, ,
	\label{np}
\end{eqnarray}
where we are supposing $\beta<-1$. For convenience, we define $E_{max}$ as
\begin{equation}
	\tau_{\gamma\gamma} (E_{max})=1 \, .
\end{equation}
It is evident from Eqs.~(\ref{tau}) and (\ref{np}) that 
$\tau_{\gamma\gamma}$ scales with energy as 
$\tau \propto E^{-(1+\beta)}$; the requirement 
$\tau_{\gamma\gamma}(100~{\rm MeV}) \gtrsim 4$ 
(see Eq.~(\ref{reqtau})) then allows us to constrain the value 
of $E_{max}$ as follows:
\begin{equation}
	1/4\gtrsim (E_{max}/100~{\rm MeV})^{-(1-3.5)} \Rightarrow E_{max}\lesssim 4^{-1/2.5} \times 100~{\rm MeV} \sim 60~{\rm MeV} \, .
	\label{emax}
\end{equation}
This requirement on $E_{max}$ implies the following condition on the Lorentz factor of the shell. Using 
$R=2 c \Gamma^2 \delta t_{obs}/(1+z)=6\times10^{10}\Gamma^2 \left[\delta t_{obs}/((1+z)\rm s)\right]$ cm 
and substituting Eqs.~(\ref{ean}) and (\ref{np}) into Eq.~(\ref{tau}) we have
\begin{eqnarray}
\Gamma\sim\left[\frac{1.8\times10^{-47}~C (d_L/{\rm cm})^2  (2.6\times10^5)^{1+\beta}}{\tau_{\gamma\gamma}(E_{max})(1+z)^{(2+2\beta)}(100)^{\beta} (\delta t_{obs}/{\rm s})(-1-\beta)(E_{max}/{\rm keV})^{(1+\beta)} }\right]^{1/(2-2\beta)} \, .
\label{gammap}
\end{eqnarray}
For interval I, setting $C=C_{Band}=584$, $\tau_{\gamma\gamma}(E_{max})=1$, 
$\beta=-3.5$, $z=0.903$ and $d_{L}=1.8\times10^{28}$ cm as 
appropriate for GRB 090510, we get 
\begin{eqnarray}
E_{max}  & \lesssim &  60~{\rm MeV} \, , \\	
\Gamma_{I}  & \sim &  160 (\delta t_{obs}/100{\rm~ms})^{-1/9}(E_{max}/60 ~{\rm MeV})^{5/18} \, .
		\label{reqns}
\end{eqnarray}

\subsubsection{Thickness for scattering on pairs}

To explain a non-thermal spectrum and a high temporal variability, 
the number of pairs created by the optically thick portion of the 
spectrum should remain small. This is in order to avoid the Thompson
optical depth for photon scattering on the created pairs becoming 
much greater than unity 
\citep[see e.g.][]{Abdo2009op,Peer2004,Guetta2001,Lithwick2003,Sari1997}. 
Since we can reasonably assume that each photon of $E>E_{max}$ 
creates a pair, the number of pairs is approximately 
\begin{equation}
	N_{pair}\sim N_{\gamma>E_{max}} \, .
\end{equation}
The Thompson optical depth is thus \citep[][]{Abdo2009op}
\begin{equation}
	\tau_{\gamma \pm} (E_{max})\sim \frac{\sigma_T N_{\gamma>E_{max}}}{4\pi R^2} \, .
\label{anc}
\end{equation}

We note that the above expression for $\tau_{\gamma \pm}$ was also 
used by \citet{Abdo2009op}, who pointed out that it's typically 
difficult to have a source optically thin to scattering on pairs 
when the optical thickness to pair production is very high. Using 
Eqs.~(\ref{np}) and (\ref{anc}), we can write
\begin{equation}
	\tau_{\gamma \pm} (E_{max}) \sim  \frac{\sigma_T C ~ (d_L/{\rm cm})^{2} (E_{max}/{\rm keV})^{1+\beta}}{4~c^2(-1-\beta)(100)^{\beta}(\delta t_{obs}/{\rm s})\Gamma^4}=2\times10^{9}(\delta t_{obs}/{100~\rm ms})^{-1}(E_{max}/60~{\rm MeV})^{-2.5}\Gamma^{-4} \, ,
\end{equation}
where we have used $\beta=-3.5$, $z=0.903$, $C=C_{Band}=584$, $d_{L}=1.8\times10^{28}$. Requiring $\tau_{\gamma \pm}\lesssim 1$ thus implies
\begin{equation}
\Gamma_I\gtrsim \Gamma_{\gamma\pm}= 200 (\delta t_{obs}/{100~\rm ms})^{-1/4}(E_{max}/60~{\rm MeV})^{-5/8} \, . \label{gammac}
\end{equation}

\subsubsection{Thickness for scattering on electrons}

Further constraints on the Lorentz factor come from scattering of 
the emitted photons on electrons inside the shell. According to the 
IS model, a fraction $\epsilon_e$ of the internal energy of shocked 
particles goes into accelerating the electrons. The shock-
accelerated electrons then radiate via synchrotron (and IC) 
emission. A necessary condition for radiation from IS to be observed 
is that the source is optically thin for photon scattering on 
electrons associated with baryons present inside the shell itself. 
When the $\tau_{\gamma e}$ for photon scattering on electrons is 
high, the spectrum of the observed radiation is modified by the 
standard assumptions of thin synchrotron and IC emission, and
effects related to the presence of the so-called electron 
photosphere need to be considered. Another condition we thus need to 
set is that 
\citep{Meszaros2005,Peer2004,Guetta2001,Lithwick2003,Meszaros2000} 
\begin{equation}
	\tau_{\gamma e}\sim \frac{\sigma_T N_{baryon}}{4\pi R^2}\lesssim 1 \, ,
\end{equation}
where we have indicated with 
\begin{equation}
	N_{baryon}=\frac{L \delta t_{obs}}{(1+z)\epsilon_e \Gamma m_p c^2}
\end{equation}
the number of electrons associated with baryons inside the shell, and 
$R \sim 2 c \Gamma^2 \delta t_{obs}/(1+z)$ is the radius of the shell. 
We thus get
\begin{equation}
	\tau_{\gamma e}\sim 10^{9} L_{52} (1+z)  (\delta t_{obs}/100 {\rm ms})^{-1} \Gamma^{-5}\epsilon_e^{-1}\lesssim 1 \, .
\end{equation}

By integrating in the $10$ keV - $100$ MeV energy range a Band 
function with normalization constant $\sim 0.08$ ph/cm$^2$/s/keV, 
$E_{peak}=2.8$ MeV, $\alpha=-0.59$ \citep{Abdo2009}, $\beta=-3.5$, 
and multiplying by $4\pi d^{2}_{L}$ (with $z=1.903$), we 
estimate the luminosity in the cosmological rest frame to be 
$L \sim 3\times10^{53}$ erg/s. Thus,
\begin{equation}
	\Gamma_{I} \gtrsim \Gamma_{\gamma e}=140~\epsilon_e^{-1/5} (\delta t_{obs}/100~{\rm ms})^{-1/5} \, .
	\label{gammaph}
\end{equation}

\subsubsection{Synchrotron emission from IS}

As we have seen in the previous sections, the spectrum observed 
during interval I could be reconciled with more commonly 
observed Band spectra by requiring the optical thickness for 
pair production to be responsible for the unusually steep 
high-energy spectral decay. Here we analyze the conditions under 
which such a spectrum could be explained as synchrotron 
emission from IS. The peak of the synchrotron component in the 
IS model is \citep{Guetta2003} 
\begin{eqnarray}
\label{piccosincr}
E_{peak}\sim 1.2\left(\frac{3p-6}{p-1}\right)^{2}
\epsilon_e^{3/2}\epsilon_B^{1/2}L_{52}^{1/2}\Gamma^{-2}(\delta t_{obs}/100~{\rm ms})^{-1}~{\rm GeV} \, .
\end{eqnarray}

For the case of GRB 090510, a high energy slope of $\beta=-3.5$ 
can be explained in the (optically thin) IS scenario by 
requiring $p\sim5$ \citep[so that the expected photon spectral 
index is $-p/2-1=-3.5$; see][]{Guetta2003} and by requiring 
that Eq.~(\ref{reqns}) is valid. Using these, and setting 
$\epsilon_e\sim \epsilon_B \sim 0.5$, $L_{52}=30$, we can write 
Eq.~(\ref{piccosincr}) as follows:
\begin{equation}
	E_{peak}=0.3(\delta t_{obs}/100~{\rm ms})^{-7/9}(E_{max}/60~{\rm MeV})^{-5/9}~{\rm MeV} \, .
\end{equation}

To have a peak around $1$ MeV, we need $\delta t_{obs} \sim 20$ ms and $E_{max}\sim 60$ MeV. For such a value of the variability timescale, from Eq.~(\ref{reqns}) we get $\Gamma_I \sim 200$, and using Eqs.~(\ref{gammac}) and (\ref{gammaph}) we also have $\Gamma_{\gamma\pm}\sim 300$ and $\Gamma_{\gamma e}\sim 200$. This implies that $\tau_{\gamma\gamma}(60~{\rm MeV})\sim1\sim\tau_{\gamma e}$, but $\tau_{\gamma \pm}\sim 6$.
We thus expect to have some effects from scattering of the 
emitted photons on the created pairs. Numerical simulations are 
the best way to predict these effects, since there are 
different processes that come into play during the dynamical 
timescale. In addition to pair production and scattering on 
electrons, discussed before, one should also consider, e.g., 
re-heating of the electron population due to synchrotron self-
absorption \citep{Ghisellini1988}, and pair annihilation. All 
these effects combined together can modify the observed 
spectrum. 

\citet{Peer2004} have carried out time-dependent numerical 
simulations within the IS model, describing cyclo-synchrotron 
emission and absorption, inverse and direct Compton scattering, 
and pair production and annihilation (including the evolution 
of high energy electromagnetic cascades), allowing a 
calculation of the spectra resulting when the scattering 
optical depth due to pairs is high, thus presenting deviations 
from the simple predictions of the thin case IS model 
\citep[e.g.][]{Guetta2003}. In particular, \citet{Peer2004} 
have shown that from moderate to large values of $\tau_{\gamma\gamma}$, 
the resulting spectrum peaks in the MeV range (as was the case 
for GRB 090510), shows steep slopes at lower energies, and 
exhibits a sharp 
cutoff at $\sim 10$ MeV. For large compactness, scattering by pairs 
becomes the dominant emission mechanism, as we have seen here for 
GRB 090510 ($\tau_{\gamma \pm} \sim 6$). 
In such a case, electrons and positrons lose their energy much 
faster than the dynamical timescale, and a quasi-Maxwellian 
distribution is formed \citep{Peer2004}. The energy gain of the 
low-energy electrons by direct Compton scattering results in a 
spectrum steeper than Maxwellian at the low-energy end, 
indicating that a steady state did not develop. Predicted 
slopes in $\nu F_{\nu}$ are $0.5\lesssim 2+\alpha \lesssim 1$ 
\citep{Peer2004}. 

In the case of GRB 090510, the low energy photon spectral slope is 
$\alpha=-0.59$. This value implies a very steep rise in the 
$\nu F_{\nu}$ spectrum, of about $\sim 1.4$. Detailed modeling of 
the spectrum for high compactness is beyond the purpose of this 
paper. These considerations, however, allow us to conclude that, 
overall, the spectrum observed during Interval I may be accommodated 
within a high compactness, synchrotron IS scenario. 


\subsection{Interval II}
\label{sec4}

\subsubsection{Transparency to GeV photons}

During interval II, photons up to 30 GeV were observed by the 
Fermi/LAT \citep{Abdo2009}. Thus, in contrast to interval I, the 
source should be optically thin to pair production and have 
$\tau_{\gamma\gamma}\lesssim 1$ at $E_{max}\gtrsim 10$ GeV. During 
Interval II, the observed spectrum is consistent with (see 
Sec.~\ref{sec1})
\begin{equation}
	N(E)=C_{pow} (E/100 {\rm~ keV})^{\beta}\frac{\rm ph}{\rm cm^{2}~s~keV} \, ,
\end{equation}
where $\beta\sim -1.58$ and $C_{pow} \sim 1.6 \times 10^{-2}$. Thus, 
using these values in Eq.~(\ref{gammap}), and setting 
$z=0.903$, $d_{L}=1.8\times10^{28}$ cm, 
$\tau_{\gamma\gamma}(E_{max})=1$, and $E_{max}= 10$ GeV, we obtain 
the following requirements:
\begin{eqnarray}
E_{max}  & \gtrsim &  10{\rm ~GeV} \, ; \\
	\Gamma_{II}  & \sim &  430  (\delta t_{obs}/100{\rm ~ms})^{-1/5.16}(E_{max}/10{\rm ~GeV})^{0.58/5.16} \, .
	\label{reqgamma2}
\end{eqnarray}


\subsubsection{Synchrotron emission from IS?}

If we assume that the spectrum observed during Interval II is 
dominated by synchrotron emission from IS; i.e., it is generated by 
the same radiation mechanism explaining the emission in Interval I, 
then it is necessary to require that the peak of the synchrotron 
component has a soft-to-hard evolution, with $E_{peak}\sim 2.8$ MeV 
during interval I and $E_{peak}\gtrsim 1 ~{\rm GeV}$ during interval 
II. Substituting Eq.~(\ref{reqgamma2}) into (\ref{piccosincr}) we have
\begin{equation}
	E_{peak}\sim 3\times10^{-5}(\delta t_{obs}/100~{\rm ms})^{-0.61}(E_{max}/10~{\rm GeV})^{-0.22}~{\rm GeV} \, ,
\end{equation}
where we have set $\epsilon_e\sim\epsilon_B\sim 0.5$, $p\sim5$.
We have also used the fact that during interval II, most of the emitted 
energy is in the GRID energy range, with a measured 25 MeV - 500 MeV 
fluence of $2.12\times 10^{-5} \rm erg~ cm^{-2}$ 
\citep{Giuliani2009}, thus giving 
$L\sim4\pi d^{2}_{L}(2.12\times10^{-5}\rm erg~ cm^{-2}/1  ~s)\sim10^{53}$ erg/s 
during interval II. From the above equation, it is evident that 
even setting $\delta t_{obs}\sim 1$ ms, we have $E_{peak}<<1$ GeV 
for $E_{max}\gtrsim10$ GeV.

\subsubsection{SSC emission from IS: a better explanation}

The extreme soft-to-hard evolution observed between interval I and 
II suggests an alternative two-component explanation. Specifically,  one could 
think of the emission in interval I being dominated by IS 
synchrotron of a slower shell with $\Gamma\sim\Gamma_{I}$ (see 
Sec.~\ref{sec3}), while the emission in interval II dominated by IS 
SSC of a late-emitted faster shell 
($\Gamma\sim\Gamma_{II}>\Gamma_{I}$), whose SSC component falls in 
the observed band, while the synchrotron counterpart is shifted to 
lower energies.

\begin{figure*}
\begin{center}
\includegraphics[width=6cm,angle=90]{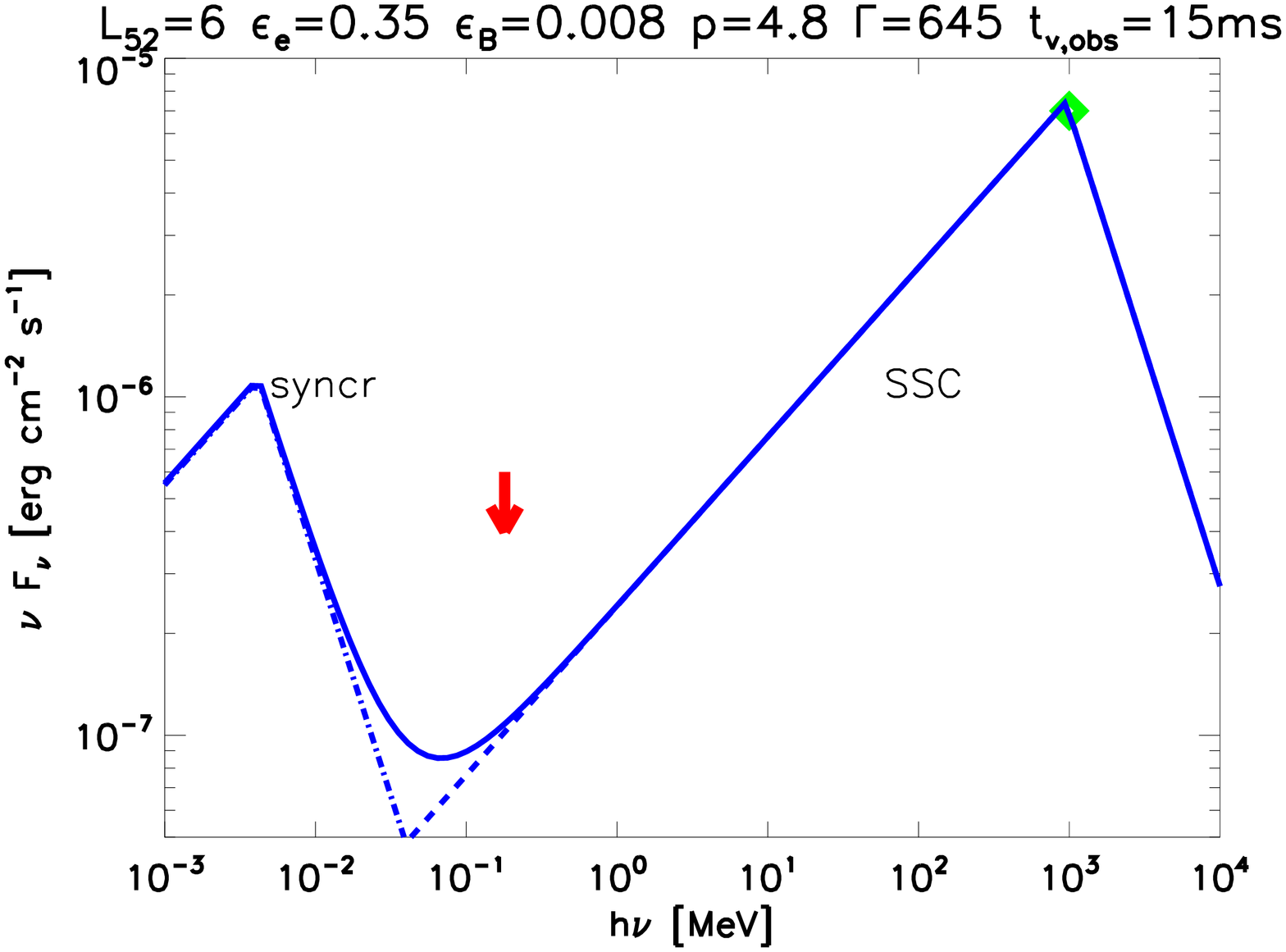}
\caption{Modeling of the high energy emission during interval II in 
the IS scenario \citep{Guetta2003}. The red arrow marks the level of 
BAT upper-limits \citep[see e.g.][]{DePasquale2009} while the green 
diamond marks the level of the flux observed by the LAT around 1 GeV 
\citep[see Fig.~2 in][]{Abdo2009}.\label{FigIS}}
\end{center}
\end{figure*}

In Fig.~\ref{FigIS} we show a possible solution within this 
scenario: during interval II, the high energy emission is dominated 
by the IC component of a faster shell with $\Gamma_{II}\sim 645$, 
$L_{52,II}\sim 6$, $\delta t_{obs,II}\sim 15$ ms, 
$\epsilon_{e,II}=0.35$, $\epsilon_{B,II}=0.008$, $p_{II}=4.8$. We 
stress that what we show in this figure implies that a viable 
parameter choice does exist to accommodate the observations within 
this model. However, such a solution is not necessarily unique, and a 
larger parameter range may exist. For a value of $p=4.8$, we expect 
a high-energy spectral slope of $\beta=-1-p/2\sim -3.4$ for the 
synchrotron component photon spectrum, consistent with our initial 
hypothesis that the true high-energy spectral slope is $\beta=-3.5$, 
and it's initially (between $T_0$ and $T_0+0.1$ s) made steeper 
($\beta\lesssim -5$) by opacity due to pair production. We note that the 
slope observed by Fermi between $T_0+0.1$ s and $T_0+0.3$ s 
\citep[$\beta=-3.09^{+0.21}_{-0.35}$,][]{Abdo2009} would agree with 
the hypothesis that we expect the spectrum to become progressively 
more transparent (see Sec.~\ref{tick}). We should, however, 
keep in mind that between $T_0+0.1$ s and $T_0+0.3$ s some effects 
due to absorption might still be present. According to the 
predictions of the IS model, we also expect a photon index of $-1.5$ 
for the SSC component, which agrees with the value of 
$\beta=-1.58^{+0.13}_{-0.11}$ observed by AGILE during interval II 
\citep{Giuliani2009}. 

We finally underline that other interesting scenarios have been 
proposed to explain the high-energy emission observed during 
interval II. For example, \citet{Wu2010} recently showed that, in the 
framework of the IS model, effects related to up-scattered 
photospheric photons may become visible, and explain the delayed 
high-energy tails observed by the Fermi/LAT. The explanation we 
propose here is thus limited to considerations based on the 
(simpler) assumption of an optically thin IS model. But, indeed, 
other scenarios are possible.

\section{Synchrotron emission from the ES}
\label{sec5}

\begin{figure}
	\centering
		\includegraphics[width=10cm]{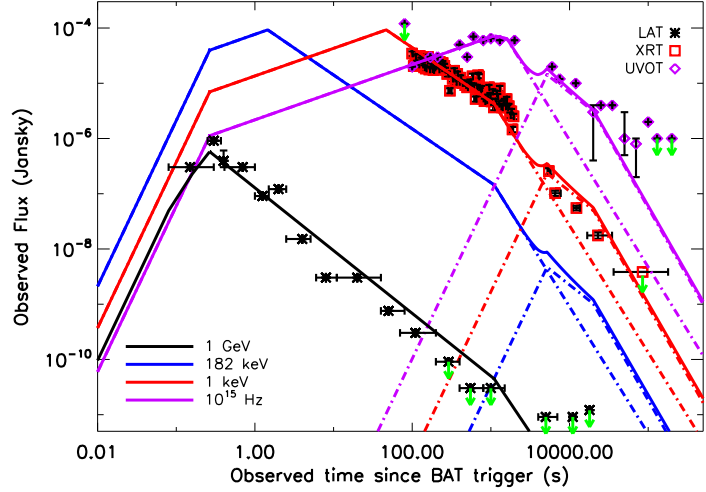}
	\caption{GRB 090510 broad-band modeling in the synchrotron ES scenario.
Data are taken from the light curves in Fig.~1 of 
\citet{DePasquale2009}, where the mean flux measured in each energy 
range for the different instruments (Fermi/LAT, Swift/XRT and 
Swift/UVOT) has been rescaled to give the measured flux at each 
specific frequency (1 GeV, 1 keV and $10^{15}$ Hz). This is done 
by requiring at 100 s the specific flux value reported in the 
SED plotted in Fig.~2 of \citet{DePasquale2009}. The black, red,
and purple solid lines represent the model predictions at 1 GeV, 1
keV and $10^{15}$ Hz, respectively. Parameters are set as follows, 
for the narrow and wide jet components (dash-dotted lines), 
respectively: $\epsilon_e=0.1$, $\epsilon_B=3\times10^{-3}$, 
$n=10^{-6}$, $\Gamma_{0,n}=10^{4}$, $E_{iso,n}=3.7\times10^{53}$ erg, 
$p_n=2.3$, $\theta_{j,n}=0.12^{\circ}$, $\Gamma_{0,w}=220$, 
$E_{iso,w}=1.5\times10^{53}$ erg, $p_w=2.5$, 
$\theta_{j,w}=0.43^{\circ}$. The blue lines represent the 
contribution of the narrow and wide jet components at the middle of 
the BAT energy band; throughout the evolution this is below the 
data/upper-limits reported in \citet{DePasquale2009}, in agreement 
with our hypothesis that the emission observed in the GBM/BAT/MCAL 
energy range should be due to IS rather than to ES.\label{Fig1}}
\end{figure}

\begin{figure}
\includegraphics[width=0.45\textwidth,height=0.25\textheight]{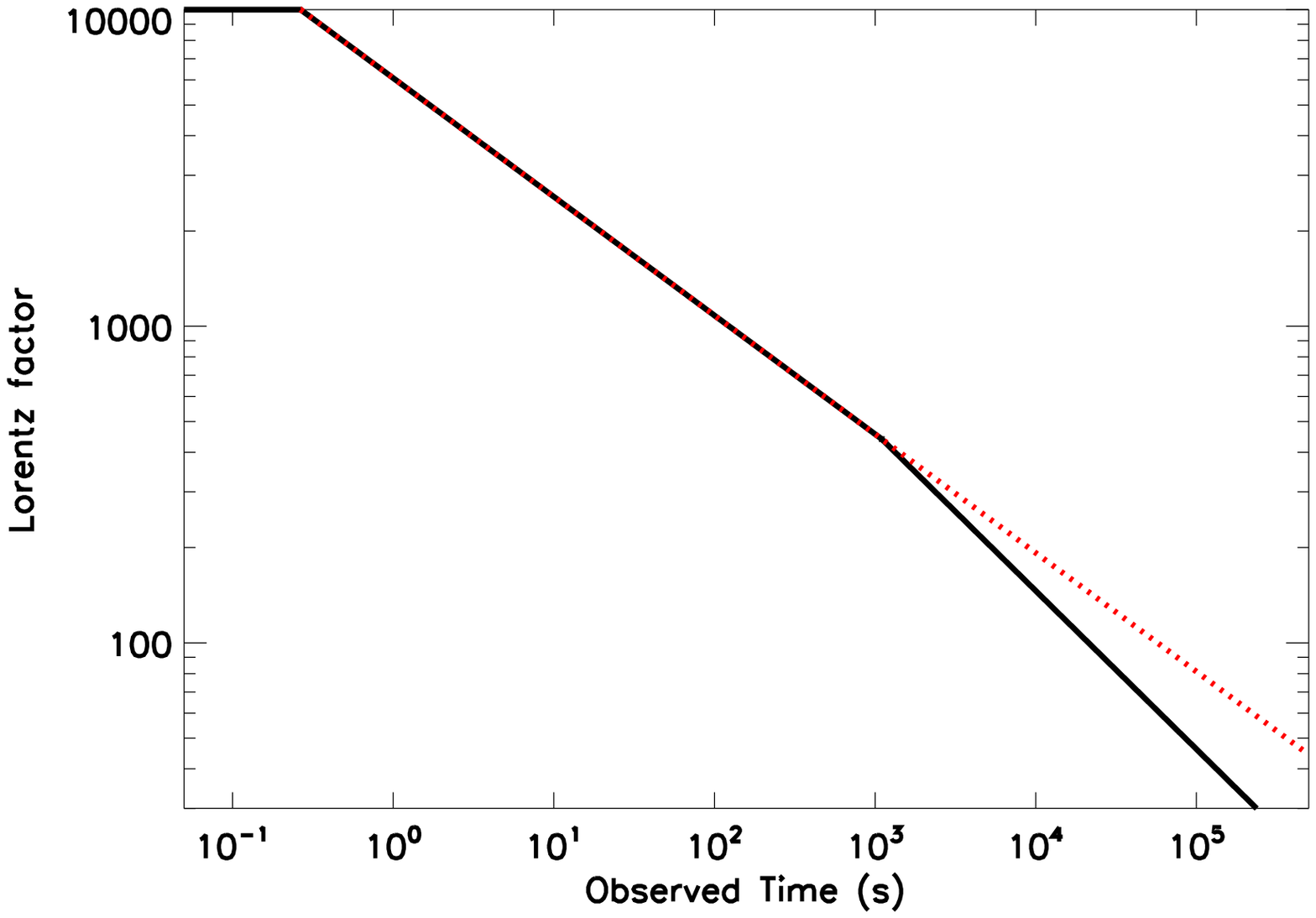}
\includegraphics[width=0.45\textwidth, height=0.25\textheight]{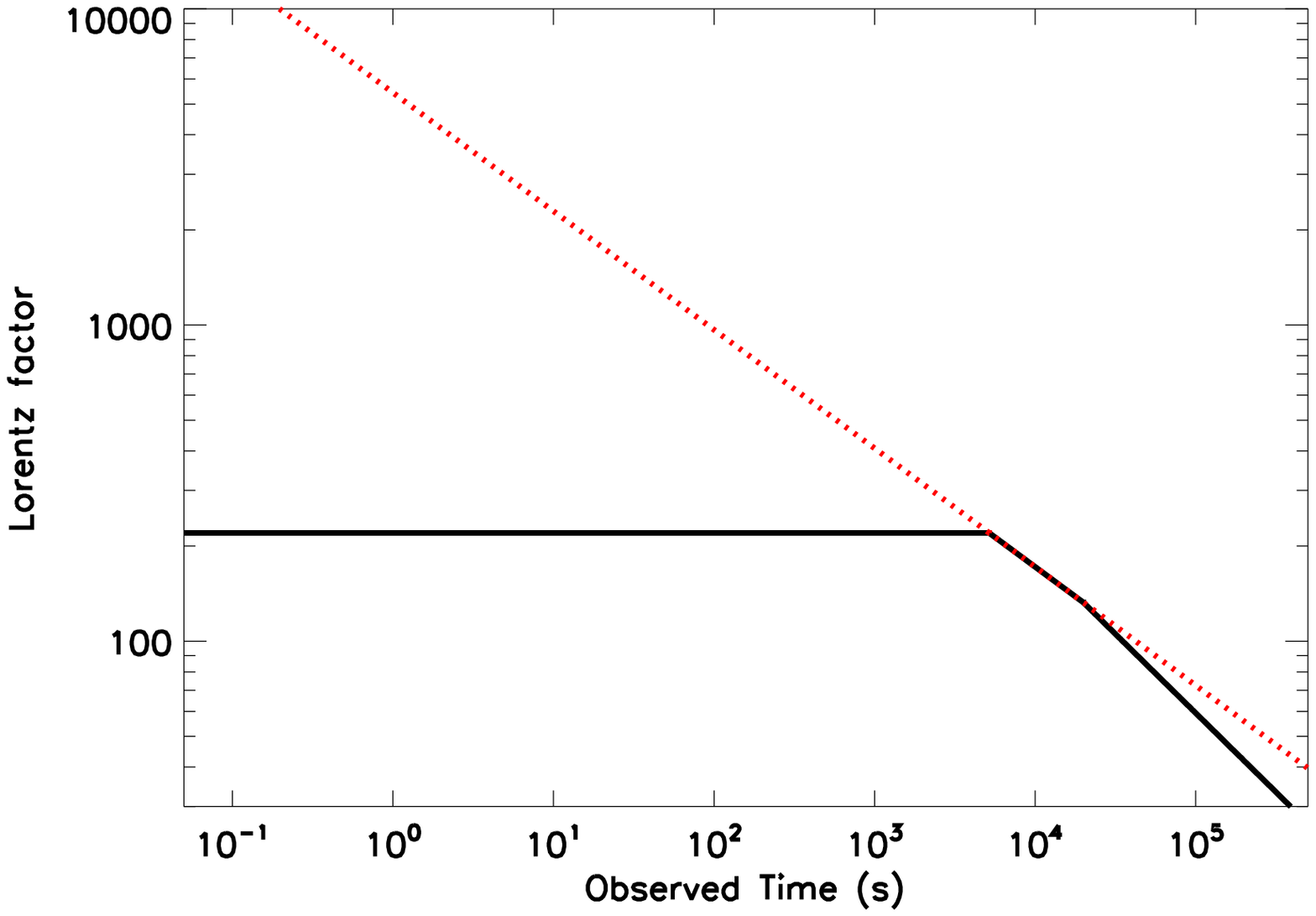}
\caption{Temporal evolution (black lines) of the Lorentz factor for 
the narrow (left panel) and wide (right panel) jet components. The 
red-dotted lines are plotted for comparison, and correspond to the 
standard evolution $\Gamma \propto t^{-3/8}$ for an adiabatic 
fireball expanding in a uniform medium \citep[e.g.][]{Sari1998}. The 
initial values $\Gamma_0$ are set to be $10^4$ and $220$, for the 
narrow and wide jet components, respectively. After $t_j$, which is 
set to be $\sim 1000$ s for the narrow jet (corresponding to 
$\theta_j \sim 0.1^{\circ}$), and $\sim 2\times10^4$ s for the wide 
jet (corresponding to $\theta_j \sim 0.4^{\circ}$), the Lorentz 
factor is evolved following a temporal scaling of 
$\Gamma \propto t^{-1/2}$ \citep[e.g.][]{Peng2005} .}
\label{gammat}
\end{figure}

In this section, we test whether the high-energy 
emission observed by the Fermi/LAT, and the optical-to-X-ray emission 
observed later on by Swift, can be explained as ES afterglow, while 
the emission in the Fermi/GBM, AGILE/MCAL and Swift/BAT is due to 
IS. In this way, one can easily account for both the high 
temporal variability observed during the prompt burst (as related to IS), and 
for the delayed onset of the high-energy emission (as related to the onset of the afterglow). This 
hypothesis, first proposed by \citet{Ghirlanda2009} on the basis of 
the temporal behavior of the high energy tail observed in the LAT up 
to 100 s after the burst, was then confirmed as a viable possibility 
by \citet{DePasquale2009} performing a broad-band analysis based on 
Swift BAT, XRT, UVOT, Fermi GBM, and LAT data. The striking feature 
of the broad-band observations is that the spectral energy 
distribution of the emission observed at 100 s is consistent with a 
single spectral component \citep{DePasquale2009}. Here we assume the 
most natural hypothesis of it being simply the synchrotron high 
energy tail\footnote{Alternatively, one could suppose that all the 
optical-to-GeV emission is generated by SSC of a synchrotron IS or 
ES component peaking at much lower energies, but this would be a 
rather non-standard scenario, which we do not analyze here.}. 

\citet{DePasquale2009} have suggested that, within the ES model, the 
peak observed around $\sim 0.2-0.3$ s after the BAT trigger in the 
Fermi/LAT light curve of the extended tail could be associated with 
the fireball deceleration time, while the peak observed in the optical range 
could be due to the synchrotron peak frequency $\nu_m$ crossing the 
band. In light of these considerations, we have modeled the 
synchrotron emission from the ES to test if a reasonable set of 
parameters does indeed exist to provide such an explanation. To this 
end, we adopt the prescriptions by \citet{Sari1998} for the 
peak flux $f_m$, the injection frequency $\nu_m$, 
and the cooling frequency $\nu_c$:

\begin{equation}
f_m\propto \Gamma^8	t^{3}_{obs} n^{3/2}\epsilon_B^{1/2}(1+z)^{-2} \, ;
\end{equation}

\begin{equation}
	\nu_m\propto
	\left(\frac{p-2}{p-1}\right)^2\Gamma^4\epsilon_e^2\epsilon_B^{1/2} n^{1/2}(1+z)^{-1} \, ;
\end{equation}

\begin{equation}
\nu_c\propto	
\Gamma^{-4}t^{-2}_{obs}n^{-3/2}\epsilon_b^{-3/2}(1+z) \, .
\label{nuc}
\end{equation}

We further rescale the expression for $\nu_c$ by a factor of 
$Y^{-2}$, to account for the effect of SSC losses on the synchrotron 
spectrum. Here $Y$ is the Compton parameter and is defined as 
follows \citep{Sari2001}:
\begin{eqnarray}
	Y & = & max\left[1, \sqrt{\frac{\epsilon_e}{\epsilon_B}}\right] ~~~~{\rm in ~fast ~cooling} \, ;\\
	Y & = & max\left[1, \left(\frac{\epsilon_e}{\epsilon_B}\right)^{1/(4-p)}\left(\frac{\nu_c}{\nu_m}\right)^{-(p-2)/2(4-p)}\right]~~~~{\rm in~slow~ cooling} \, ,
\end{eqnarray}
where $\nu_c$ is the non-rescaled value of the cooling frequency, as 
in \citet{Sari1998} and Eq.~(\ref{nuc}). To model the behavior of 
the high-energy tail, we consider the whole evolution of the Lorentz 
factor $\Gamma$ of the shell, using an approximate sharp transition 
from the coasting phase, when 
\begin{equation}
	\Gamma\sim\Gamma_0,
\end{equation}
to the deceleration phase, when \citep[e.g.][]{Sari1998} 
\begin{equation}
	\Gamma(t_{obs})=\Gamma_0(t_{obs}/t_{dec})^{-3/8} \, .
	\label{gammaevol} 
\end{equation}
Here $t_{dec}$ is the deceleration time in the observer's frame, 
given by \citep{Sari1999}
\begin{equation}
	t_{dec}=\left(\frac{3 E_{iso}}{32\pi\Gamma_0^8 n m_p c^5}\right)^{1/3}(1+z) \, ,
\end{equation}
where $m_p$ is the proton mass.

In Fig.~\ref{Fig1} we show what we obtain for the parameter choice 
$\Gamma_0=10^{4}$, $n=10^{-6}$, $\epsilon_B=3\times10^{-3}$, 
$\epsilon_e=0.1$, $E_{iso}=3.7\times10^{53}$ erg. These values are 
consistent with the results by \citet{Kumar2009}, in which the GeV 
light curve was modeled starting at $\sim 1$ s after the BAT 
trigger, while here we are modeling also its peak around 0.2-0.3 s. 
The high value of the Lorentz factor is required to have 
$t_{dec}\sim 0.3$ s, so as to explain the peak observed in the LAT 
light curve. We also note that while the very low density value is 
still consistent with those that can be expected around short GRBs 
in the coalescing binary progenitor scenario \citep[see e.g.][]{Bel2006}, 
the isotropic energy is much higher (though comparable to the one 
derived from the fluence observed in the LAT, see \citet{Abdo2009}). 

In the hypothesis that the steepening observed in the XRT light 
curve is due to a jet break, we have further evolved the Lorentz 
factor as $\Gamma \propto (t_{obs}/t_j)^{-1/2}$ \citep[see e.g.][and Fig. 3]{Peng2005}, 
finding that $t_j\sim 1000$ s in good agreement with the data. Using 
the relation \citep[e.g.][]{Rhoads1997}
\begin{equation}
	\Gamma(t_{j})=1/\theta_j \, ,
\end{equation}
and considering Eq.~(\ref{gammaevol}), we can constrain the jet 
opening angle to be
\begin{equation}
	\theta_j=\Gamma_0^{-1}(10^{3}~{\rm s}/0.3~{\rm s})^{3/8}\sim0.1^{\circ} \, .
\end{equation}
The energy in the jet is thus
\begin{equation}
	E_{j}=(\theta_j^2/2) E_{iso} \sim 7\times 10^{47}~{\rm erg} \, ,
\end{equation}
which is more easily explained in a binary merger model. 

We note, however, that after the X-ray break the optical flux 
decreases with a slope shallower than the X-ray one. 
\citet{DePasquale2009} and \citet{Kumar2009} have suggested that 
this may be explained by a jet break made shallower from the passage 
of $\nu_m$ through the optical band. With our choice of parameters, 
$\nu_m$ is crossing the optical band around the jet break time, and 
the light curve decay is still too steep (at least using our simple 
approximation of the Lorentz factor evolution). We therefore test 
the alternative hypothesis of a two-component jet, with a narrow jet 
component explaining the early time emission, and a wider component 
contributing at late times to explain the excess observed in the 
optical band. For example, \citet{Peng2005} considered such a model to 
explain the optical light curve of GRB 030329. By assuming for both 
jet components the same $\epsilon_e$, $\epsilon_B$ and $p$, 
\citet{Peng2005} found that the addition of a wider, slower 
component with $\Gamma_{0,w}\sim (1/10) \Gamma_{0,n}$ and 
$E_{j,w}\sim 4 E_{j,n}$ could explain the late-time optical excess 
observed in the light curve. A structured jet model has also been 
invoked in other cases \citep[e.g.][]{Racusin2008} to explain 
chromatic jet breaks. 

In light of these considerations, we have attempted to explain 
the optical excess observed in the case of GRB 090510 after $t_{obs}\sim 10^{3}$ s 
by adding the contribution of a wider jet component. We find that 
the choice $E_{w,iso}=1.5\times10^{53}$, $\Gamma_{0,w}\sim (1/45) \Gamma_{0,n}=220$, 
$\theta_{j,w}=0.4^{\circ}$, and $p=2.5$ (with the other parameters 
left unchanged) can account for the excess observed in the optical, 
with little contribution in the X-rays (see Fig. 2 and Fig. 3).  This choice also implies 
$E_{j,w}\sim 6 E_{j,n}$. The chosen value of $p=2.5$ is 
larger than the one adopted for the narrow component. This is 
motivated by the fact that the X-ray decay observed before $10^3$ s, 
dominated by the narrow component, is shallower than the one 
observed after $10^3$ s in the optical band 
($\delta=-1.13^{+0.17}_{-0.09}$), which we model as the emission 
from the wider component. With $p=2.5$, for $\nu_m<\nu_{opt}<\nu_c$ 
one gets a predicted value of the temporal decay index of 
$-3/4(p-1)\sim -1.12$, in agreement with the observed one within the 
uncertainties. Incidentally, we note that for the case of GRB 080319B, 
\citet{Racusin2008} obtained different $p$ values for the 
narrow- and wide-jet components, as we are finding here.

We finally test whether, for our choice of parameters, the 
contribution of SSC emission to the observed flux is indeed 
negligible (as suggested by the SED at 100 s being consistent with a 
single spectral component; see \citet{DePasquale2009}). The peak 
flux of the SSC component, in the Thomson limit, is related to the 
synchrotron one by $f^{IC}_{m}\sim10^{-6}n (R/10^{19}{\rm cm})f_m$ 
\citep{Sari2001}. In our case, the peak of the synchrotron component 
is constrained to fit the optical flux measured by the UVOT, which 
is about $10^{-4}$ Jy (see Fig.~\ref{Fig1}). This means that for 
$n=10^{-6}$, the SSC component has a flux level below $10^{-16}$ Jy 
at all energies throughout the evolution, so that its 
contribution to the light curves plotted in Fig.~\ref{Fig1} is 
completely negligible.

\section{Conclusion}
\label{sec6}
We have analyzed GRB 090510 in the context of the synchrotron IS and 
ES scenarios. We first attempted to explain the soft-to-hard 
spectral evolution associated to the delayed onset of a GeV tail with 
the hypothesis that both the prompt burst and the high-energy tail 
originate from synchrotron emission of electrons accelerated by IS. 
Considerations of the compactness of the source lead us to conclude 
that the high-energy tail should be produced in IS developing in a 
late-emitted shell, characterized by a Lorentz factor of the order 
of $\Gamma \sim 700$, greater than the one generating the prompt 
burst ($\Gamma \sim 200$). However, this condition on the Lorentz 
factor implies a hard-to-soft evolution of the peak frequency of the 
IS synchrotron component, which does not agree with the observed 
soft-to-hard evolution. 

Given the difficulties of explaining the prompt and delayed high 
energy emission with a single mechanism (synchrotron emission from 
IS), we then tested two double-component scenarios. In the first, 
the emission observed during interval I is explained as synchrotron 
emission from IS, while the high-energy tail observed in interval II 
is explained as SSC emission from IS. In the second scenario, the high 
energy emission observed by the LAT is decoupled from the prompt 
burst, and has an external shock origin. This last scenario has the 
advantage of explaining in a simple way the smooth temporal behavior 
of the high-energy tail, up to 100 s after the burst, and the 
consistency of the broad-band SED observed at 100 s with a single 
spectral component. In the ES scenario, we show that a reasonable 
set of parameters does indeed exist to explain the optical-to-GeV 
observations of this burst, despite a high Lorentz factor being 
required to have the fireball entering the deceleration phase as 
early as $t_{obs}\sim 0.3$ s (when the emission in the LAT is 
observed to peak).  The ES scenario thus seems to account more 
naturally for the observations, even if a more detailed modeling of 
the late-time chromatic break is required. We have suggested that a 
structured jet may indeed be a viable explanation of such a 
chromatic feature.

In conclusion, we stress that the high Lorentz factor implied by the 
ES scenario has some relevant consequences in relation to the 
physics of the central engine. The commonly accepted fireball model 
invokes a series of shells expanding outward with Lorentz factor of 
the order of a few hundred, where IS first generate the prompt 
emission, and then, with the merged shell continuing to expand 
outward toward the external medium, an ES generates the afterglow. 
If, on the other hand, the high-energy tail is attributed to the ES 
emission, this implies that the source emits first a very fast 
shell, which impacts on the external medium creating an early 
afterglow, plus a series of slower shells that catch up with each 
other generating the prompt $\gamma$-ray emission. At the end of the 
IS phase, the merged, slower shell (with a more typical Lorentz 
factor of the order of a few hundred), would also decelerate and 
eventually generate an afterglow by interaction with the external 
medium. In this respect, we note that the additional component 
required to explain the shallow decay observed at late times in the 
optical band could be related to the ES generated by such a 
slower shell.

\acknowledgments
The authors are grateful to Patrick Sutton for helping improve the manuscript by carefully proofreading its final version. 
A. C. thanks the Italian L'Or\'eal-UNESCO program ``For 
Women in Science'', and EGO- European Gravitational Wave 
Observatory, for support. The authors also acknowledge the support of 
ASI-INAF contract I/088/06/0.

\bibliography{Cors0706}

\end{document}